\documentclass[12pt]{iopart}
\usepackage{subfigure}
\usepackage{tabularx}
\usepackage{graphicx}	
\usepackage{color}
\usepackage{subeqnarray}
\usepackage{multirow}
\usepackage{booktabs}
\usepackage[sort&compress,numbers]{natbib}

\begin{document}

\title[Antenna for the detection of audio-band electromagnetic disturbances in LISA]{Antenna for the detection of electromagnetic audio-band disturbances on-board LISA}



\author{D~Serrano$^{1,2}$, A~Pérez-Ortega$^{1,2}$,
D~Roma-Dollase$^{1,2}$, J~Salvans-Tort$^{1,2}$, J~Ho-Zhang$^{1,2}$,  J~Ramos-Castro$^{2,3}$ and M~Nofrarias$^{1,2}$}

\address{$^1$ Institut de Ci\`encies de l'Espai (ICE, CSIC), 08193 Cerdanyola del Vall\`es, Spain}
\address{$^2$ Institut d'Estudis Espacials de Catalunya (IEEC), 08860 Castelldefels, Spain}
\address{$^3$ Department d'Enginyeria Electr\`onica, Universitat Polit\`ecnica de Catalunya,  08034 Barcelona, Spain}
\begin{abstract}

The LISA mission will be the first observatory to detect gravitational waves from space within the millihertz frequency band. Magnetic forces have an important impact on the instrument's sensitivity below the millihertz. Hence, monitoring the magnetic environment within each of the LISA spacecrafts is of utmost importance. In this Letter we present the characterization of the coils that were used in LISA Pathfinder (LPF) when operating as magnetic sensors in the audio frequency band. The necessity of implementing this type of magnetometer is presented in order to monitor high frequency magnetic signals from the electronics on-board. We show that the LPF coils have a performance one order of magnitude better than the current requirements set by the LISA mission at the low end of the audio-band frequency. The LPF coils are able to measure a magnetic noise level of 1.45 $\rm pT/\sqrt{\rm{Hz}}$ at 50 Hz and 0.17 $\rm pT/\sqrt{\rm{Hz}}$ at 500 Hz. Additionally, the LPF coils can reach a magnetic noise floor of 0.1 $\rm pT/\sqrt{\rm{Hz}}$ at frequencies above 1 kHz.

\end{abstract}

\maketitle


\section{Introduction} \label{sec:intro}

LISA is the future space-borne gravitational wave detector that aims to unveil the gravitational wave spectra in the sub-Hz band~\citep{amaroseoane2017laserinterferometerspaceantenna}, a frequency regime unattainable for ground based observatories.
The mission concept consists of three spacecraft in heliocentric orbits trailing the Earth by 20 degrees. The constellation will form an equilateral triangle of approximately 2.5 million km arm length, each spacecraft hosting two masses in nominal free fall. In such a configuration, gravitational waves interacting with the LISA constellation will produce differential displacements between the spacecraft-to-spacecraft test masses, which can be detected through heterodyne laser interferometry down to picometer-level sensitivity.

As in on-ground gravitational wave detectors, building the instrument noise model is fundamental in achieving the instrument nominal performance. An exhaustive study on the LISA noise model was carried out by the precursor mission LISA Pathfinder (LPF)~\citep{Anza05, Antonucci12} which successfully tested the technology needed for gravitational wave detection in space~\citep{Armano16, Armano18}. Among the different studies performed, LPF provided a wealth of information in terms of the local environmental conditions, in particular the temperature~\citep{Armano19_Temp}, magnetic~\citep{Armano20_Mag, Armano25a, Armano25b} and radiation~\citep{Armano18_GCR1, Armano18_GCR2} environment for future gravitational wave detectors. 

The LPF magnetic diagnostics subsystem comprised four triaxial fluxgate magnetometers and two sets of injection coils. Measurements from the magnetometers enabled the separation of the magnetic field contributions originating from the interplanetary magnetic field (IMF) and those generated by the spacecraft's on-board electronics~\citep{Armano20_Mag}. The coils, meanwhile, were used to apply controlled magnetic forces to the free-falling test masses (TMs). In combination with the magnetometers, these experiments helped characterize the magnetic properties of the TMs~\citep{Armano25a}, revealing that the dominant magnetic contribution to the instrument's overall noise budget arose from low-frequency fluctuations in the IMF~\citep{Armano25b}.

As in its precursor mission, LISA will carry a set of precision sensors with the objective of carefully monitoring any disturbances with the potential to disturb the main scientific measurement. The so-called LISA Science Diagnostics Subsystem (SDS) follows closely the design of the diagnostics subsystem in LISA Pathfinder~\citep{Sanjuan07, DiazAguilo13, Canizares09, Canizares11} adding some improvements on the temperature sensors~\citep{Roma23}, the radiation monitor~\citep{Mazzanti23} and the magnetometers~\citep{MATEOS2018311}. Since no magnetically induced force experiments are planned for LISA, this opened the possibility of repurposing the LPF injection coils as electromagnetic sensors operating in the audio frequency band ---a frequency range that was not covered by the fluxgate magnetometers aboard LPF.

In this contribution, we characterize the LISA Pathfinder coils as detectors of magnetic signals in the audio frequency band, aligning with the baseline currently being implemented in LISA. The significance of employing a sensor capable of detecting signals across this frequency range ---whilst LISA frequency band lies within the millihertz--- will be discussed. Additionally, we study the coil's performance under a controlled laboratory environment, reducing Earth's magnetic contribution to better simulate the space environment where they will operate.

This work is organized as follows. In section~\ref{sec:Mag_force}, we describe the motivation behind the implementation of audio frequency band magnetic sensors on-board the LISA spacecrafts. In section~\ref{sec:SCS}, we introduce search coil sensors (SCS) as the most commonly used detectors for this frequency band. Section~\ref{sec:LPF_coil_theory} presents the expected performance of the LPF coil when used as a SCS, which we experimentally verify in Section~\ref{sec:LPF_coil_results}. Section~\ref{sec:det_range} introduces the expected detection range of the LPF coils and compares it with an initial mock model of the LISA spacecraft's size and disposition. Finally, the main conclusions are presented in Section~\ref{sec:conclusions}.
\section{Audio-Band Antenna Motivation} \label{sec:Mag_force}

LISA operation frequency band lies within the millihertz range, however, higher frequency magnetic fields have a potential to induce forces to the LISA test masses. Different mechanisms could be considered to account for this contribution. In~\citep{Armano18} it was proposed that part of the excess noise measured in the LISA Pathfinder low frequency band could be originated in the down-conversion of high frequency magnetic fields through the $B\cdot\nabla B$ product on the magnetic-induced force expression. Another possibility to consider is that a given high frequency magnetic field has a slow variable amplitude modulation that couples into the millihertz band.

It is worth noting that the vacuum enclosure and the electrode housing surrounding the TMs are expected to shield them from high frequency magnetic fields, with optimal effectiveness above the kilohertz range. However, the test mass shielding may be not so effective for frequencies around the few hundreds hertzs (at the low end of the audio frequency band, ranging from 20 Hz to 20 kHz)~\cite{prediction_for_LPF}, which could turn into spurious effects measured in the test mass. Having an electromagnetic sensor to monitor the audio-band could therefore be useful during in-flight operations to raise quality flags, in case any disturbances are detected. Such a sensor on-board LISA must be designed to meet the following sensitivity requirement

\begin{equation}
  S_{\rm  B, req}^{1/2} < \sqrt{1 + \frac{10^{6}}{f^{2}}} \,\rm{pT}/\sqrt{\rm{Hz}}, \quad \rm{50\, Hz} < f < \rm{500\, Hz }.
\label{eq:LISA_requirement}
 \end{equation}

In the following sections, we provide further details on the derivation of high-frequency forces acting on the test mass, as well as on the experimental validation of the proposed antenna design.

\subsection{Magnetic force modulations}\label{sec:Dipole_approx}

LISA test masses can be approximated as behaving like magnetic dipoles.  The magnetic force they feel under the influence of an external magnetic field, $\mathbf{B}$, can then be described as

\begin{equation}
    \mathbf{F} = \left\langle \left(\mathbf{m}\cdot\mathbf{\nabla}\right)\mathbf{B} \right\rangle V,
    \label{eq:Magnetic_force}
\end{equation}

where $\mathbf{m}$ is the total magnetic moment density of the TM and $\left\langle \ldots\ \right\rangle \equiv \frac{1}{V}\int \,(\ldots)\,d^{3}x$ denotes the average of the enclosed quantity over the TM volume, $V$. The TM total magnetic moment density has two components, the remanent magnetic moment density, $\mathbf{m}_{r}$, and the induced magnetic moment density, $\mathbf{m}_{i}$. The latter depends linearly on the external magnetic field, for diamagnetic and paramagnetic materials, and can be expressed as

\begin{equation}
    \mathbf{m}_{i} = \frac{\chi}{\mu_0}\mathbf{B},
    \label{eq:induced_moment}
\end{equation}

where $\chi$ is the magnetic susceptibility and $\mu_{0}$ the vacuum permeability. Because of the test masses composition, 73\% gold and 27\% platinum, we can safely consider a diamagnetic response to external magnetic fields as shown in~\cite{Armano25a}. In terms of magnetic susceptibility, the worst case scenario will take place for a magnetic susceptibility of $\chi = -1$, which would represent the TM repelling the external magnetic field, exerting the largest force possible. 

Taking all of this into account, the magnetic force from~(\ref{eq:Magnetic_force}) becomes

\begin{equation}
    \mathbf{F} = \left(\mathbf{M}_{r}\cdot\mathbf{\nabla}\right)\mathbf{B} + \frac{\chi V}{\mu_{0}}\left( \mathbf{B}\cdot\nabla\right)\mathbf{B},
    \label{eq:Magnetic_force_total}
\end{equation}

where $\mathbf{M}_{r}=\mathbf{m}_{r}V$ is the remanent magnetic moment, and we have assumed an isotropic and homogeneous magnetic susceptibility for the TMs. 
The external magnetic field $\mathbf{B}$, for this analysis, will originate from the AC magnetic sources within the spacecrafts, which we will approximate as magnetic dipoles of negligible volume. Hence, the magnetic field they generate at a given distance $\mathbf{r}$ can be described as

\begin{equation}
    \mathbf{B_{\rm dipole}} = \frac{\mu_0}{4\pi} \left[ \frac{3 ( \mathbf{p} \cdot \mathbf{r} ) \mathbf{r}}{r^5} - \frac{\mathbf{p}}{r^3} \right],
    \label{eq:dipole_b_field}
\end{equation}

where $\mathbf{p}$ is the magnetic moment of the dipole and $r=||\mathbf{r}||$. For simplicity, and without loss of generality, we will consider the worst case scenario where the magnetic field amplitude is maximum, which is the case for $\mathbf{p}$ and $\mathbf{r}$ being aligned. In that situation, the dipole magnetic field from~(\ref{eq:dipole_b_field}) becomes

\begin{equation}
    \mathbf{B_{p}} = \frac{\mu_0}{2\pi r^{3}}\mathbf{p}.
    \label{eq:dipole_b_field_worst_case}
\end{equation}

This is the magnetic field in the stationary case. However, we are interested to see the effect of a frequency dependent current. We will consider that the sources magnetic dipolar behavior is originated by a localized system of currents that oscillate sinusoidally with time such that 

\begin{equation}
    \mathbf{J}(\mathbf{r}, t) = \mathbf{J}(\mathbf{r})e^{-i\omega t},
    \label{eq:J_with_time}
\end{equation}

where $\mathbf{J}$ is the current density and $\omega = 2\pi f$ is the angular frequency of the oscillations at a frequency $f$. Using Ampère's law, the magnetic field produced by such distribution of currents is

\begin{equation}
        \mathbf{B} = \frac{\mu_0}{4\pi} \bigl\lbrace  k^2 (\mathbf{n} \times \mathbf{p}) \times \mathbf{n} \frac{e^{i\omega t}}{r}  + \left[ 3 (\mathbf{p} \cdot \mathbf{n})\mathbf{n} - \mathbf{p} \right] \left( \frac{1}{r^3} - \frac{ik}{r^2} \right) e^{i\omega t} \bigr\rbrace,
    \label{eq:B_dipole_with_frequency}
\end{equation}

where $k=\omega/c$, $r=ct$ and $\mathbf{n}=\mathbf{r}/r$. Considering that the sources will lie within the dimensions of the spacecrafts, the distances will never be larger than 1 m to the TMs. Additionally, the audio-band frequency is also small enough to assume that $r\ll c/\omega$. Thus, the dominant component of the magnetic field from~(\ref{eq:B_dipole_with_frequency}), at the locations of the TMs, will be the term proportional to $1/r^{3}$. By comparing with the static case from~(\ref{eq:dipole_b_field}), the magnetic field can be expressed as an oscillating magnetic dipole

\begin{equation}
    \mathbf{B} = \mathbf{B_{p}}e^{i\omega t},
    \label{eq:B_near_zone}
\end{equation}

which we have assumed to be in the maximum amplitude orientation, as shown in~(\ref{eq:dipole_b_field_worst_case}). This type of magnetic fields will not induce forces within the frequency band of LISA. However, the amplitude of the currents may have a small variation with time, originated from stationary random processes, that can oscillate more slowly, acting as an envelope for the high frequency magnetic fields. In order to introduce such amplitude modulations we will consider an additional term into~(\ref{eq:B_near_zone}) such that

\begin{equation}
    \mathbf{B} = A(t)\,\mathbf{B_{p}}e^{i\omega t},
    \label{eq:B_coils_paper}
\end{equation}

where $A(t)$ is adimensional ---more details on this parameter in the following. Now that we have derived the high frequency magnetic field from the electronics of the spacecraft, the magnetic force acting on the TMs can be estimated. Without loss of generality, we will work with the $x$ component and consider both $\mathbf{p}$ and $\mathbf{r}$ in the $x$ direction. Introducing~(\ref{eq:B_coils_paper}) into~(\ref{eq:Magnetic_force_total}), we obtain

\begin{equation}
    F_x = -3M_{x}A(t)\frac{B_{0}}{r} - 3A^{2}(t)\frac{\chi V}{\mu_0}\frac{B_{0}^{2}}{r},
    \label{eq:Magnetic_force_final}
\end{equation}

where we have defined

\begin{equation}
    B_{0} \equiv |\mathbf{B_{p}}|e^{i\omega t},
    \label{eq:Bx0}
\end{equation}

as the worst case scenario amplitude of the magnetic field dipole sinusoidal signal.


In order to numerically evaluate the previous, we will assume small amplitude fluctuations of $A(t)$, so that we can expand $A(t)\approx 1 + \delta A$ and, by neglecting second order terms in~(\ref{eq:Magnetic_force_final}), derive the force amplitude spectral density (ASD), $S^{1/2}$, to be

\begin{equation}
    S^{1/2}_{F_{x}} = 3\frac{B_{0}}{r}\left( M_{x} + 2\frac{\chi V}{\mu_{0}}B_{0} \right)S^{1/2}_{A}.
    \label{eq:ASD}
\end{equation}

To put some numbers into perspective, we will consider the following values for the parameters involved: $M_{x} = 0.14$ nAm\textsuperscript{2}~\cite{Armano25a}, $r = 0.3$ m, a value for the side length of the TMs cube of $L = 46$ mm, $B_{0} = B_{\rm rms}\sqrt{2} = \sqrt{2}$ nT, a percentage of  $20\%$ modulation of $A(t)$ and a worst case scenario for the magnetic susceptibility of the TMs of $|\chi| = 1$. The latter assumption follows the magnetic susceptibility model derived in~\cite{susceptibility_aprox}, verified at low frequencies by~\cite{Armano25a}, which predicts that at around 450 Hz, the TMs magnetic susceptibility starts to ramp up towards $\chi = -1$ of the superconductive infinite frequency limit. Taking all of this into account, the acceleration noise, $S^{1/2}_{a} = S^{1/2}_{F_{x}}/m_{TM}$, with a value of the TMs mass of $m_{TM} = 1.96$ kg, is estimated to be around $3.17\times10^{-16}$ ms$^{-2}$Hz$^{-1/2}$. Under such conditions, the estimated contribution is below the measured $(1.74\pm0.05)\times10^{-15}$ ms$^{-2}$Hz$^{-1/2}$ above 2 mHz from LPF main results~\cite{Armano18}, representing approximately a  potential 3\% contribution in noise power. 

Our estimate of the high frequency magnetic fields contribution to acceleration noise would increase if we relax the small modulation condition and impose that $A(t)$ behaves, for instance, as a square wave. This would  increase the estimate of the acceleration noise contribution by typically one order of magnitude~\cite{large_amplitude}, hence becoming a potential dominant source of acceleration noise in the TMs. Such a case is not to be considered a nominal one since units on-board LISA have stringent requirements to prevent this scenario. Nevertheless, this situation further reinforces the rationale for implementing a magnetic sensor capable of monitoring high-frequency magnetic fields originating from electronic units, as a precaution in the event of non-nominal conditions. 

\section{Search Coil Sensors} \label{sec:SCS}

Search coil magnetometers—consisting of thousands of turns of thin wire, often wound around a ferromagnetic core to enhance performance—have played a central role in the study of electromagnetic waves in space plasmas. Their wide frequency range, high sensitivity, compact size, low mass, and low power consumption make them ideal for spaceflight applications. The first magnetometer flown by the United States was a SCS aboard Pioneer 1~\citep{5008730}. Since then, they have been widely used in numerous missions for planetary exploration and space weather monitoring, as detailed in~\citep{Coils_missions} and references therein. Air-core search coil magnetometers are reported to be at their fundamental sensitivity limits~\citep{s21165568} unlike their ferromagnetic counterpart, which continue to be investigated --such as in the study presented in~\citep{2024_ferromagnetic}.

Amongst many other magnetic sensors, SCS have the largest amplitude detection range, down to around 1 pT and without an upper limit to their sensitivity~\cite{56910}. This type of sensors most useful detection frequency band lies within the 1 Hz - 1 MHz, can have diameters from 1 cm to 100 cm, with a mass on the order of 100 grams and require between 1 and 10 mW of power to operate the companion electronics. The principle behind SCS is Faraday's induction law, a changing magnetic flux through a coiled conductor induces a voltage proportional to the changing rate. In the case of ferromagnetic SCS, a high permeability material is placed within the core of the wire winding in order to increase its performance as it focuses magnetic field lines into the coil surface. 

The sensitivity of SCS magnetometers, defined as the ratio of the induced voltage in the coil to the external magnetic field, depends on the permeability of the core material, the coil area, the number of turns and the magnetic flux rate of change through the coil. The frequency response of coil sensors is limited by the ratio of their auto-inductance ($L$) to their resistance ($R$) ---this ratio sets the upper limit of the detection frequency band. The higher the value of $L$, the slower the dissipation of the induced current within the coil once the external field is removed. Additionally, when multiple turns are involved, the coil own windings can be considered as an effective arrangement of capacitors storing energy. Hence, the interwinding self-capacitance of the coil ($C$) will also limit the frequency response.

Under ideal circumstances the sensitivity of coil magnetometers can be directly derived from Faraday's induction law such that

\begin{equation}
    S_{0} \equiv \frac{\rm V}{B} = N\omega \pi a^{2},
    \label{eq:sensitivity_search_coil_simple}
\end{equation}

where $\rm V$ is the induced voltage, $B$ is the external magnetic field, $N$ is the number of turns, $a$ is the radius of the coil. As we can see from~(\ref{eq:sensitivity_search_coil_simple}), the sensitivity can either be improved by increasing the radius, the number of turns or both. Ferromagnetic SCS main advantage is that the coil dimensions can be kept relatively small because an additional $\mu_{r}$ factor appears within the ideal sensitivity, which represents the relative permeability of the ferromagnetic material. The downside is that the sensitivity will no longer be linear in the ideal case due to demagnetization factors coming into play. Moreover, the ferromagnetic core would increase the low frequency noise of the sensor due to thermal fluctuations.

In LISA, it is of key importance to minimize magnetic field gradients within the TMs locations, hence precluding the ferromagnetic core SCS. We consider in our study an air-core SCS, despite the better performance of their ferromagnetic counterpart. 

\subsection{Equivalent circuit}

\begin{figure}
    \centering
    \includegraphics[width=\linewidth]{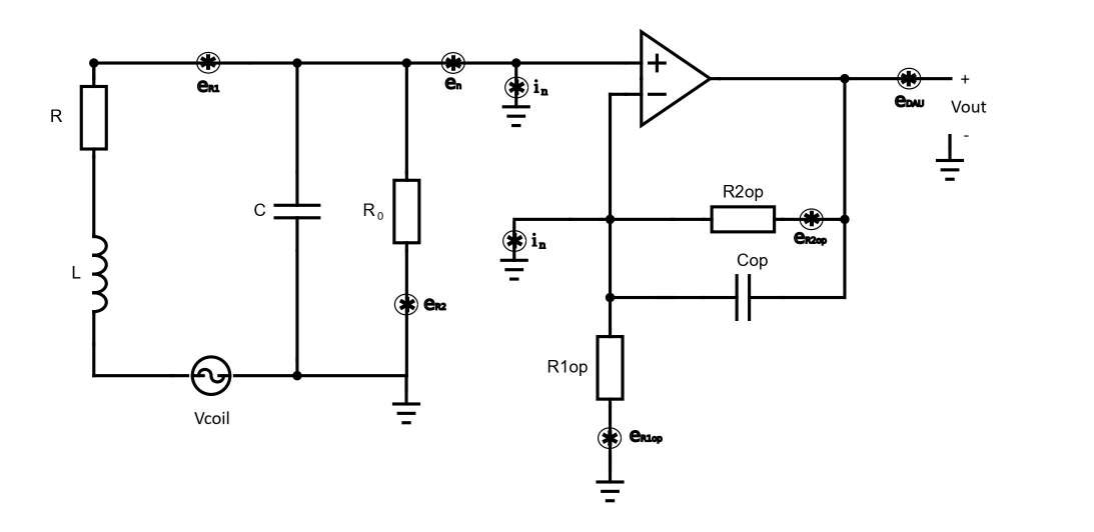}
    \caption{Equivalent circuit of the LPF coil used as a SCS (left) and its amplification circuit (right). The noise sources are also presented as circled stars.}
    \label{fig:Search_coil_circuit}
\end{figure}

The coil's idealized response is not really representative since, in reality, we have to also consider autoinductance and self-capacitance effects and not only its resistance. On the leftmost part of figure~\ref{fig:Search_coil_circuit}, we show the equivalent RLC circuit of any coil when employed as a SCS. We have included a load resistor, $R_{0}$, in parallel to the coil equivalent circuit. Its main purpose is to fine-tune the sensor's sensitivity smoothness with frequency and to prevent the coil's typical resonant frequency characteristic. The signal provided at the output of the coil is generally weak, which is the reason why an operational (OP) amplifier needs to be connected to the sensor to amplify the signal. In order to minimize the amplifier own noise from dominating the overall noise of the system it requires a high accuracy of amplification and a very low voltage noise and flat response in the measuring bandwidth. The amplifier topology is that of a non-inverter with a total gain of

\begin{equation}
    G = 1 + \frac{R_{2op}}{R_{1op}\left(1 + i\omega R_{2op}C_{op}\right)},
    \label{eq:amplifier_gain_coil}
\end{equation}

where $R_{1op}$ and $R_{2op}$ are the resistors of the OP amplifier and an extra capacitor, $C_{op}$, is also set in parallel to $R_{2op}$ acting as a low-pass filter. 

\subsection{Sensitivity}

Now that we have described the circuit that will be employed for the LPF coils, when used as a SCS, we can derive a formula for its sensitivity. To do so, we will first calculate the transfer function of the coil itself as

\begin{equation}
    H_{R} = \frac{V_{out}}{V_{in}} = \frac{1}{1 + \frac{Z_{1}}{Z_{2}}},
    \label{eq:transfer_function_coil}
\end{equation}

which relates the induced voltage in the coil, $\rm V_{in}$, due to external magnetic fields, with the output voltage, $\rm V_{out}$, that would appear at the terminals of the LPF coil prior to the amplification process. $Z_{1}$ is the equivalent impedance of the inductor, $L$, and the resistor, $R$, and $Z_{2}$ is the one of the capacitor, $C$, and the load resistor, $R_{0}$. The fraction of the impedances can be expressed as

\begin{equation}
    \frac{Z_{1}}{Z_{2}} = \frac{R}{R_{0}} - \omega^{2}LC + i\omega\left( RC + \frac{L}{R_{0}} \right).
    \label{eq:impedances_fraction}
\end{equation}

Combining equations~(\ref{eq:transfer_function_coil}) and~(\ref{eq:impedances_fraction}), the transfer function can be described as

\begin{equation}
    H_{R} = \frac{1}{1 + \frac{R}{R_0} - \omega^{2}LC + i\omega\left( R C + \frac{L}{R_0} \right)}.
    \label{eq:transfer_func_modulus}
\end{equation}

The final sensitivity of the SCS circuit is related to the idealized sensitivity from~(\ref{eq:sensitivity_search_coil_simple}) as $S = |H_{R}|S_{0}$ such that

\begin{equation}
    S =\frac{N\omega \pi a^{2}}{\left[(1+\alpha)^2+\gamma^2\left(\frac{\alpha^2}{\beta^2}+\beta^2-2\right)+\gamma^4\right]^{\frac{1}{2}}},
    \label{eq:SCS_sensitivity}
\end{equation}

where

\begin{equation}
    \alpha \equiv \frac{R}{R_0}, \quad \beta \equiv R \sqrt{C / L}, \quad \gamma=\frac{\omega}{\omega_0},
    \label{eq:alpha_beta_gamma_transf_func}
\end{equation}

where $\omega_0 = 2\pi f_{0} = 1/\sqrt{L C}$ is the resonance angular frequency of the sensor. The sensitivity response increases linearly, initially, with frequency up to the resonance frequency. Above $\omega_0$ the influence of the self-capacitance of the coil causes the output signal to drop and, with it, the sensitivity of the SCS.

\subsection{Noise analysis}\label{sec:intro_noise}

In the schematic shown in figure~\ref{fig:Search_coil_circuit}, we also display the expected sources of noise that will contribute to the noise measurement of the SCS, marked as stars. To determine the equivalent magnetic noise of the sensor, the equivalent voltage noise at the amplifier input has to be characterized. We will derive it by taking into account all the circuit equivalent noise sources: the amplifier voltage and current noise, resistors thermal noise and the data acquisition unit (DAU) noise. Each of the elements will have their own transfer functions. 

For $R_{0}$ the transfer function corresponds to the configuration in series of $R$, $L$ in parallel with $C$ and $R_{0}$

\begin{equation}
    H_{R_0} = \frac{1} {\frac{1}{R + i\omega L} + i\omega C + \frac{1}{R_0}} \cdot \frac{1}{R_0},
    \label{eq:H_r2}
\end{equation}

For resistor $R_{1op}$, the open loop gain of the OP amplifier has to be taken into account

\begin{equation}
    A_{0} = \frac{\omega_{0}}{\omega_{0} + i\omega}A_{\nu 0},
    \label{eq:loop_gain}
\end{equation}

with $\omega = 2\pi f_{op}$ and $A_{\nu 0}$ the signal voltage gain of the OP amplifier. Thus,

\begin{equation}
    H_{R_{1op}} = \frac{\frac{R_{2op}}{1 + R_{2op} \cdot i\omega C_{op}} \cdot A_{0}}{R_{1op}(A_{0} + 1) + \frac{R_{2op}}{1 + R_{2op} \cdot i\omega C_{op}}}.
    \label{eq:H_r1op}
\end{equation}

Similarly for $R_{2op}$

\begin{equation}
    H_{R_{2op}} = \frac{A_{0}\left(1 + i\omega R_{2op}C_{op}\right)}{1 + A_{0} + \frac{R_{2op}}{R_{1op}}}.
    \label{eq:H_r2op}
\end{equation}

Finally, for the overall amplifier

\begin{equation}
    H_T = \frac{A_{0}}{1 + \frac{R_{1op}}{R_{1op} + \frac{R_{2op}}{1 + R_{2op} \cdot i\omega C_{op}}} A_{0}}.
    \label{eq:H_T}
\end{equation}

Now that we have computed all the transfer functions of all components we can estimate the total noise budget of the sensor. Firstly, the voltage noise density contribution from the DAU is assumed to be white noise within the frequency regime of interest and estimated to be

\begin{equation}
    e_{DAU} = \frac{\sigma_{rms}}{G\sqrt{f_{n}}},
    \label{eq:e_DAU}
\end{equation}

where $f_{n}$ is the Nyquist frequency which is half of the sampling frequency, $f_{n} = f_{s}/2$, and $\sigma_{\rm rms}$ is the RMS white noise of the DAU. Then, we also have the noise contributions coming from the OP amplifier. The amplifier equivalent voltage noise density can be modelled as a function of frequency

\begin{equation}
    e_{op} = e_{0}\frac{H_{T}}{G}\sqrt{1+\frac{f_{\nu c}}{f}},
    \label{eq:OP_noise}
\end{equation}

where $f_{\nu c}$ is the corner frequency of the voltage noise density for the OP amplifier and $e_{0}$ is the voltage noise density as provided by the product specifications. The amplifier current noise density for the non-inverting input is also coupled into voltage noise by the equivalent coil impedance such that

\begin{equation}
    e_{in} = \frac{H_{T}}{G}i_{0}Z\sqrt{1+\frac{f_{i c}}{f}},
    \label{eq:OP_current_noise}
\end{equation}

where $f_{i c}$ is the corner frequency of the current noise density for the OP amplifier and $i_{0}$ is the current noise density. Additionally, the current noise due to the inverting input of the operational amplifier generates an equivalent noise cuh that

\begin{equation}
    e_{in2} = \frac{i_{0}}{G}\left( \frac{R_{2op}}{1 + R_{2op}i\omega C} \right)\sqrt{1+\frac{f_{i c}}{f}}.
    \label{eq:OP_own_current_noise}
\end{equation}

All resistances thermal noise densities have to be included into the total noise model as well

\begin{equation}
    e_{R} = e_{B}H_{R}\frac{H_T}{G},
    \label{eq:e_R1}
\end{equation}

\begin{equation}
    e_{R_0} = e_{B}H_{R_0}\frac{H_T}{G},
    \label{eq:e_R2}
\end{equation}

\begin{equation}
    e_{R_{1op}} = e_{B}\frac{H_{R_{1op}}}{G},
    \label{eq:e_R1op}
\end{equation}

\begin{equation}
    e_{R_{2op}} = e_{B}\frac{H_{R_{2op}}}{G},
    \label{eq:e_R2op}
\end{equation}

where $e_{B} = \sqrt{4K_{B}TR_i}$ is the thermal noise density of a resistor with resistance $R_i$, at temperature $T$ and with $K_{B}$ being the Boltzmann constant. 

Lastly, by assuming that all the different noise contributions are uncorrelated, the total noise budget can be expressed as the sum in quadrature of all of the previously derived terms

\begin{equation}
        e_{T} = \bigl(e_{DAU}^2 + e_{op}^2 + e_{in}^2 + e_{in2}^2 + e_{R}^2 + e_{R_{0}}^2 + e_{R1op}^2 + e_{R2op}^2\bigr)^{1/2}.
    \label{eq:noise_total}
\end{equation}
\section{LPF Coil Sensor Characterization \label{sec:LPF_coil_theory}}

To keep the heritage from LPF, we chose to reuse the coil design employed as a magnetic signal injector in the precursor mission, repurposing it as an audio-band sensor for LISA. The LPF coil consisted of 2400 turns of copper wire winding assembled on a titanium alloy frame, $\rm Ti_{6}\, Al_{4}$, with an average radius of 56.5 mm. Although our primary focus is the audio-band regime, we present results across the full operational range of typical SCS ---1 Hz to 1 MHz--- as introduced in Section~\ref{sec:SCS}.

\begin{table}
    \centering
    \caption{Key parameters specifications of all components involved in the circuit from figure~\ref{fig:Search_coil_circuit} for the LPF coil. Values obtained experimentally are presented with their corresponding uncertainties.}
    \begin{tabular}{cc}
        Parameter & Value \\ \midrule\midrule
        $R$ [$\Omega$] & $1814.2 \pm 0.1$ \\
        $L$ [H] & $1.355 \pm 0.002$  \\ 
        $C$ [pF] & $404.7 \pm 0.1$  \\
        $R_{0}$ [$\Omega$]  & 7500 \\
        $G$ & 101  \\
        $f_{c}$ [kHz] & 250  \\
        $C_{op}$ [pF] & 15  \\
        $f_{ic}$ [Hz] &  3 \\
        $f_{\nu c}$ [Hz] & 0.3 \\
        $f_{op}$ [Hz] & 1.2 \\
        $R_{1op}$ [$\Omega$] & 250  \\
        $R_{2op}$ [k$\Omega$] & 25  \\
        $A_{\nu 0}$ ($\times 10^{6}$) & 3.1 \\
        $e_{0}$ [nV$/\sqrt{\rm{Hz}}$] & 6.9  \\
        $i_{0}$ [pA$/\sqrt{\rm{Hz}}$] & 0.2  \\
        $\sigma_{RMS}$ [$\mu$V]& 13  \\
    \end{tabular}
    \label{tab:coil_params}
\end{table}

The key parameters obtained for the LPF coil are listed within table~\ref{tab:coil_params}. The first set of parameters, $R$, $L$ and $C$, correspond to the LPF coil structure and were measured experimentally, hence the uncertainties are provided. The coil resistance, $R$, was measured using a digit multimeter, 34470A from Keysight. The coil self-capacitance, $C$, and autoinductance, $L$, were measured using an electronic impedance analyzer, IM3590 from Hioki, which provided the coil's impedance and phase as a function of frequency. From the maximum phase and its frequency the autoinductance can be derived as $L = Z_{L}/2\pi f$. The self-capacitance can be derived from the minimum phase of the impedance and its frequency as $C = 1/(2\pi f Z_{C})$.

A load resistance, $R_{0}$, is placed in parallel to the RLC circuit of the LPF coil, see the leftmost part of the circuit in figure~\ref{fig:Search_coil_circuit}. The implementation of this load resistance into the SCS is one of the typically employed methods to improve the sensor frequency characteristic response~\cite{Tumanski_2007}. For low values of the load resistance, one is able to operate in the so-called self-integration mode, which is the plateau between the low corner frequency 

\begin{equation}
    \omega_{l} = 2\pi f_{l} = \frac{R + R_0}{L}.
    \label{eq:low_corner}
\end{equation}

and the high corner frequency

\begin{equation}
    \omega_{h} = 2\pi f_{h} = \frac{1}{R_0 C}.
    \label{eq:high_corner}
\end{equation}

\begin{figure}
    \centering
    \includegraphics[width=\linewidth]{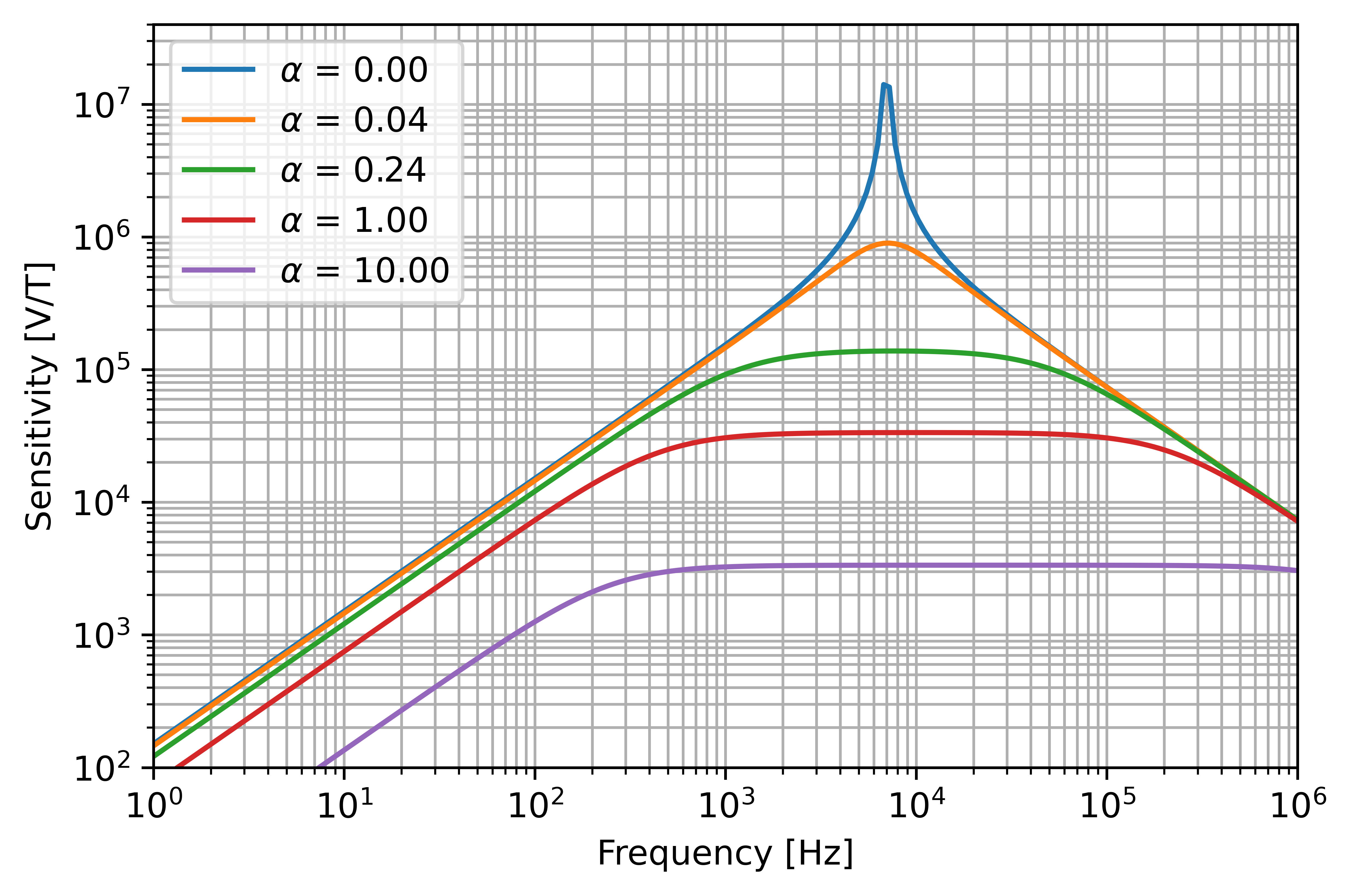}
    \caption{LPF coil sensitivity as a function of frequency for different values of the load resistance, $\alpha = R/R_{0}$.}
    \label{fig:R2_dependency}
\end{figure}

$R_{0}$ can be considered a free parameter that we can use in our favor to achieve a sufficient smooth response of the antenna. In order to select a suitable value, we build the SCS sensitivity as a function of frequency using~(\ref{eq:SCS_sensitivity}) for different values of the ratio $R$ to $R_{0}$. As we show in figure~\ref{fig:R2_dependency}, the sensitivity frequency response flattens for increasing values of $\alpha$, a reduction of the load resistance, for frequencies around the resonance frequency of the LPF coil, $f_0 \approx 7$ kHz, and in the range limited by $f_{l}$ and $f_{h}$. We aim to achieve the highest possible sensitivity while avoiding resonances within the sensor, which requires a frequency response that is as flat as possible. The value chosen for the load resistance of our system, $R_{0} = 7500 \, \Omega$ (green line in figure~\ref{fig:R2_dependency}), is such that the trade-off of removing the resonance of the coil at frequencies around the kHz does not reduce much the sensitivity and its linearity within the frequency band of interest, 50 - 500 Hz.

The amplifier selected for the circuit was an operational amplifier ADA4077 from Analog Devices. We defined a DC gain of 101 in order to guarantee a good enough performance for the amplifier by selecting the values of $R_{1op}$ and $R_{2op}$, following~(\ref{eq:amplifier_gain_coil}). The amplifier capacitance, $C_{op}$, was chosen allowing a frequency cutoff of $f_{c} = 250$ kHz, acting as a low-pass filter. The acquisition unit for the voltage measurements selected was a NI-DAQ USB-6212, from National Instruments. The main reason to chose this DAU was its high sampling frequency reaching up to $f_{s} = 400$ kHz, allowing for a very high frequency characterization of the SCS. Additionally, its very low random noise contribution, $\sigma_{RMS}$, when operating at the lowest nominal full scale range,implied a negligible contribution into the final equivalent noise density of entire sensor. All the key parameters required for the noise estimation of both devices can be obtained from their specifications and we included them in Table~\ref{tab:coil_params}. 



\begin{figure}[t]
    \centering
    \includegraphics[width=\linewidth]{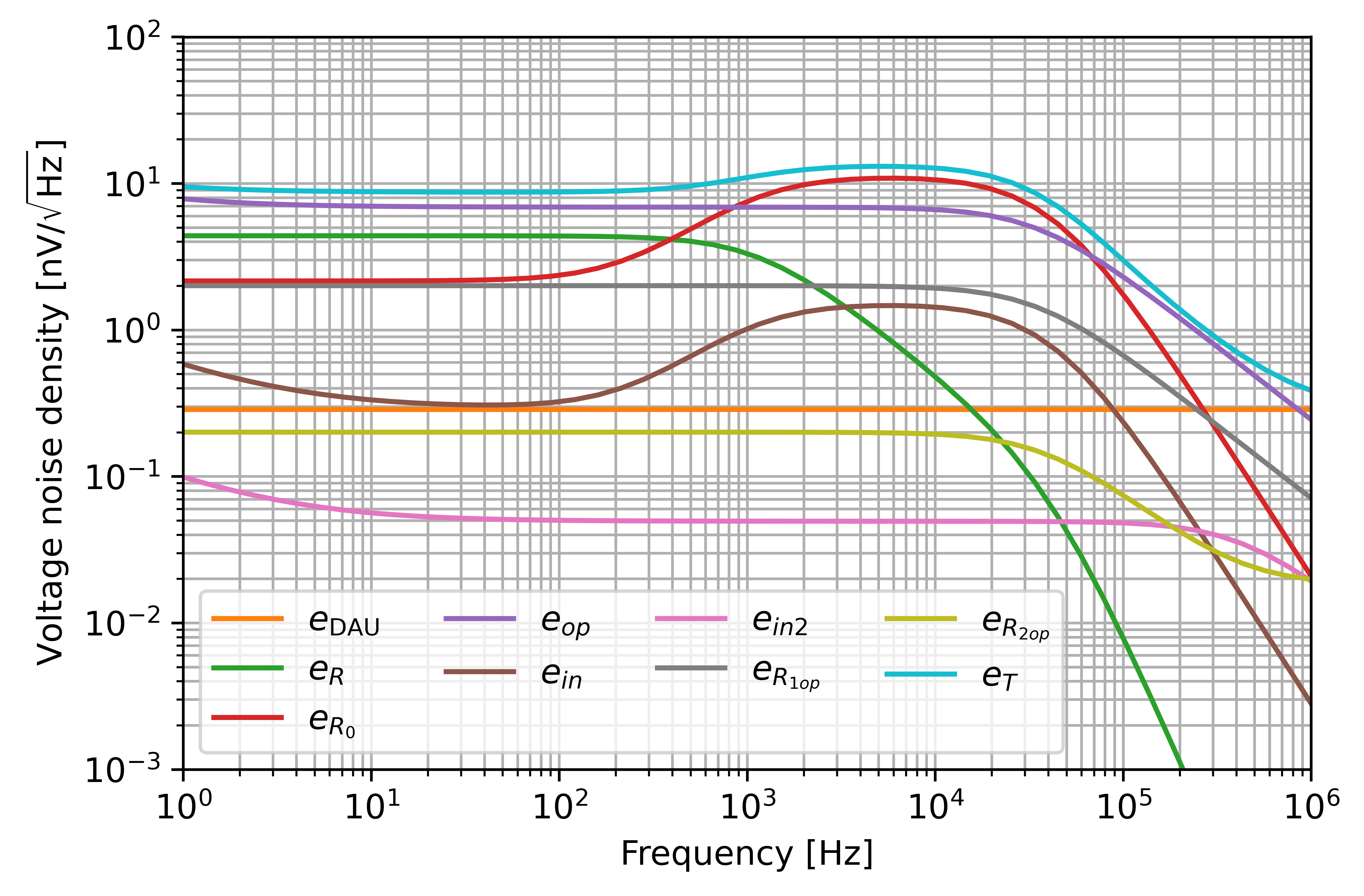}
    \caption{LPF coil's equivalent voltage noise amplitude spectrum density. We include all the noise contributions from all components involved in the circuit from figure~\ref{fig:Search_coil_circuit}.}
    \label{fig:LPF_coil_noise}
\end{figure}

Taking into account all sources of noise involved within the SCS circuit from figure~\ref{fig:Search_coil_circuit} we can build the predicted equivalent voltage noise for our coil magnetometer. Using  equations in Sec.~\ref{sec:intro_noise} together with the parameters presented in table~\ref{tab:coil_params}, we obtain the curve from figure~\ref{fig:LPF_coil_noise}. We can observe the equivalent voltage noise contributions of the different components at the amplifier output as a function of frequency. The selected dynamic range of the DAU, as previously mentioned, provides a negligible white noise level up until the MHz. The noise level is thus dominated by the amplifier noise, $e_{op}$, and the thermal noise from the LPF coil, $e_{R}$, for frequencies below 0.1 kHz. At higher frequencies, the thermal noise of the LPF SCS decays and becomes negligible but the load resistance noise, $e_{R_{0}}$, starts dominating alongside the amplifier. The current system with the operational amplifier selected is already close to the limit of the LPF coil intrinsic noise levels limited by its structural design. 


Finally, we can obtain the equivalent magnetic noise curve by dividing the equivalent voltage noise output by the sensitivity curve in~(\ref{eq:sensitivity_search_coil_simple}) at the corresponding frequencies

\begin{equation}
    e_{B} = \frac{e_{T}}{S}.
    \label{eq:B_noise}
\end{equation}

\begin{figure}
    \centering
    \includegraphics[width=\linewidth]{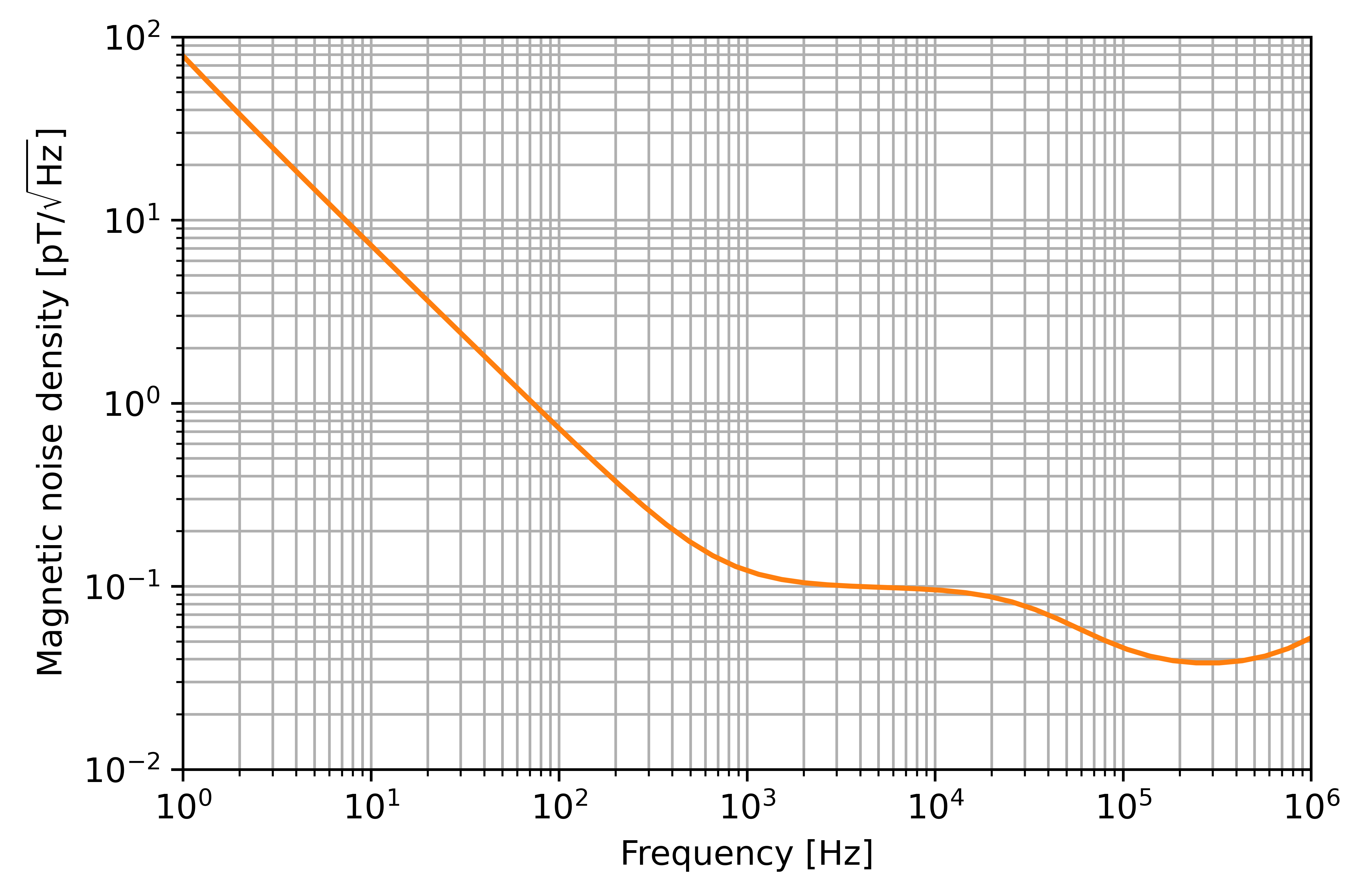}
    \caption{Amplitude spectrum density of the LPF coil's magnetic noise level.}
    \label{fig:B_noise_curve}
\end{figure}

We display the magnetic noise density in figure~\ref{fig:B_noise_curve}, where we are no longer showing each of the individual noise contributions. The results of our performance analysis show that the LPF coils can achieve a noise floor down to 0.1 pT/$\sqrt{\rm{Hz}}$ at frequencies between 1 kHz and 0.1 MHz, even without the use of a ferromagnetic core. Within the audio frequency band we the LPF coil is expected to reach 1.45 pT/$\sqrt{\rm{Hz}}$ at 50 Hz and 0.17 pT/$\sqrt{\rm{Hz}}$ at 500 Hz.

\section{Experimental Results \label{sec:LPF_coil_results}}


In order to evaluate the coil performance we will require the use of a triaxial Helmholtz coil able to induce controlled and homogeneous magnetic fields within the entire LPF coil surface. A three layer mu-metal shield will also be employed to suppress the environmental magnetic field, dominated by Earth's magnetic field and by the building's electricity mains. The shield geometry consists of three flat ended cylinders of decreasing radius one placed inside the other. On one side all flat ended caps have a small 22 mm diameter hole for power cables to pass through. Additionally, we make use of a TFM100G4-S fluxgate magnetometer from Billingsley to get a calibrated measurement of the magnetic field injected by the Helmholtz coil.


First, we induce a set of sinusoidal signals through the Helmholtz coil at frequencies rangin from 5 Hz to 1 kHz. The axis of injection is aligned with the axis of the LPF coil for maximum output. To guarantee an optimal alignment, we built a dedicated structure and placed it within the Helmholtz coil. The advantages provided were twofold, it guaranteed that the alignment between the coils axes was correct within a $\pm 0.1^{\circ}$ and it also ensured that the LPF coil was placed within the spherical region of homogeneous magnetic field of around 150 mm in diameter, to ensure that differences in magnitude remain below 1\%.  

\begin{figure}
    \centering
    \includegraphics[width=\linewidth]{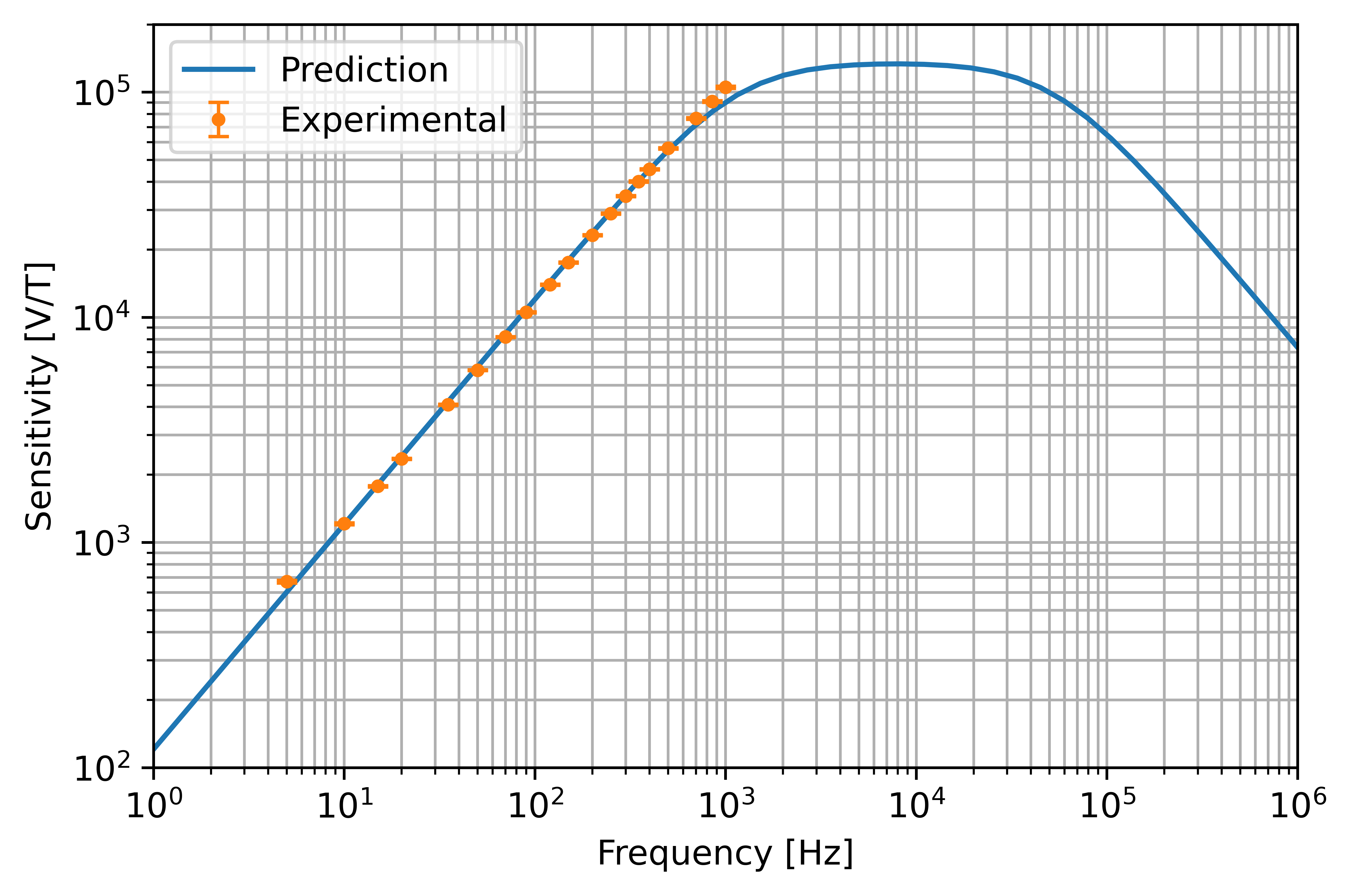}
    \caption{LPF coil sensitivity prediction alongside experimental results.}
    \label{fig:sensitivity_with_data}
\end{figure}

We did not apply magnetic shielding to the LPF coil during the sensitivity estimation runs. The reason being that the injected signals were large enough in magnitude such that external contributions were negligible in comparison. The fluxgate magnetometer was placed in the same location of the coil for the same signals but in different runs. We decided not to have both the fluxgate and the LPF coil at the same time since fluxgate magnetometers have high permeability ferromagnetic cores, which could affect the performance of the coil. The values of the magnetic field measured from the fluxgate magnetometer together with the voltage output of the LPF coil allowed the determination of the sensitivity, $S = V/B$. The results of the estimated sensitivity show good agreement with the predicted sensitivity, as illustrated in figure~\ref{fig:sensitivity_with_data}. Above 900 Hz, however, we observe a noticeable discrepancy, which we attribute to the limited operational frequency range of the fluxgate sensor ---approximately 1 kHz.

Once verified the sensitivity model of the SCS circuit, we proceed to estimate the sensor's noise performance. To do so the LPF coil is placed inside the mumetal chamber to isolate the system from external interference. Typically, within the lab, there are two major sources of external magnetic fields. Firstly, Earth's magnetic field dominates the offset (DC) of the measurements with values reaching up to 60 $\mu$T, depending on the latitude and radial distance from its center. The mumetal shield suppresses this contribution by up to three orders of magnitude, depending on proper alignment of the axis with respect to Earth's magnetic field component. Secondly, we also want to remove the lab's own sources of magnetic fields. These are typically dominating at high frequencies and originate primarily from the power lines of the building at around 50 Hz and its harmonics. Proper grounding of the mumetal chamber is mandatory, mainly of its innermost layer, to reduce the high frequency noise from capacitive coupling into the system. 

\begin{figure}
    \centering
    \includegraphics[width=\linewidth]{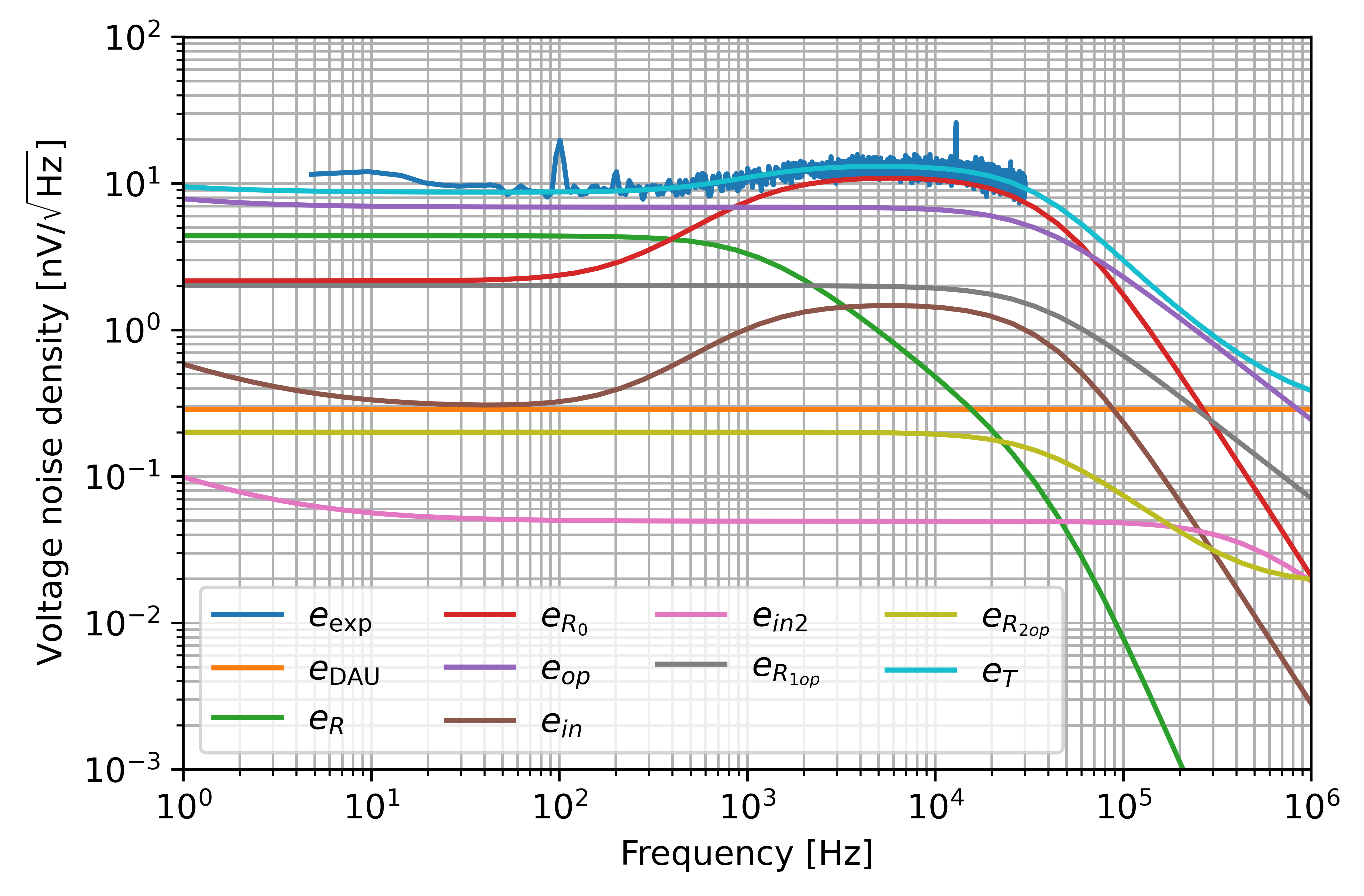}
    \caption{LPF coil equivalent voltage noise density together with experimental results obtained for a measurement run of 5 seconds. The first 2 data points, at low frequencies have been removed to prevent window biasing. Frequencies larger than 30 kHz have also been removed due to coupling into the power cables feeding the coil.}
    \label{fig:voltage_with_data}
\end{figure}

The resulting measured voltage noise for the LPF coil measurements are shown in figure~\ref{fig:voltage_with_data}. The data consisted on a 5 seconds run at the maximum sampling frequency of 400 kHz. To determine the equivalent voltage noise density of the coil we calculated the ASD of the measurements using Welch's method with a Blackmann-Harris window function with a segment length of 4\% of the entire sample length with a 50\% overlap. We can observe that the experimental results match the prediction model from Eq.~\ref{eq:noise_total}. However, we want to note a couple of undesired signals at 100 Hz and at 12.8 kHz. The origin of such signals is unknown but irrelevant for this study (the former may be originated from the full bridge rectification of 50 Hz mains and the latter is outside of our band of interest). These are typically  signals coupling into the mumetal shield from other equipment connected to the power mains in the building at the same time as the data was taken. In reality, these spikes do not affect the performance of the SCS since, they are originated from external sources. If anything, these spikes prove that the LPF sensor is good at measuring low amplitude oscillating magnetic fields. Additionally, the ASD at frequencies higher than 30 kHz was removed from the final plot. The main reason why we removed them was that they were coupling into the cables from the power supply, outside the chamber, into the LPF coil inside the mumetal shield. The only possible way to remove these signals would have been to change the power supply by batteries, which would fit inside the shielding, preventing external high-frequency coupling into the cables since they would be inside the mumetal. Nonetheless, we did not consider taking this approach since the audio frequency range of interest is already showing the predicted performance.

\begin{figure}
    \centering
    \includegraphics[width=\linewidth]{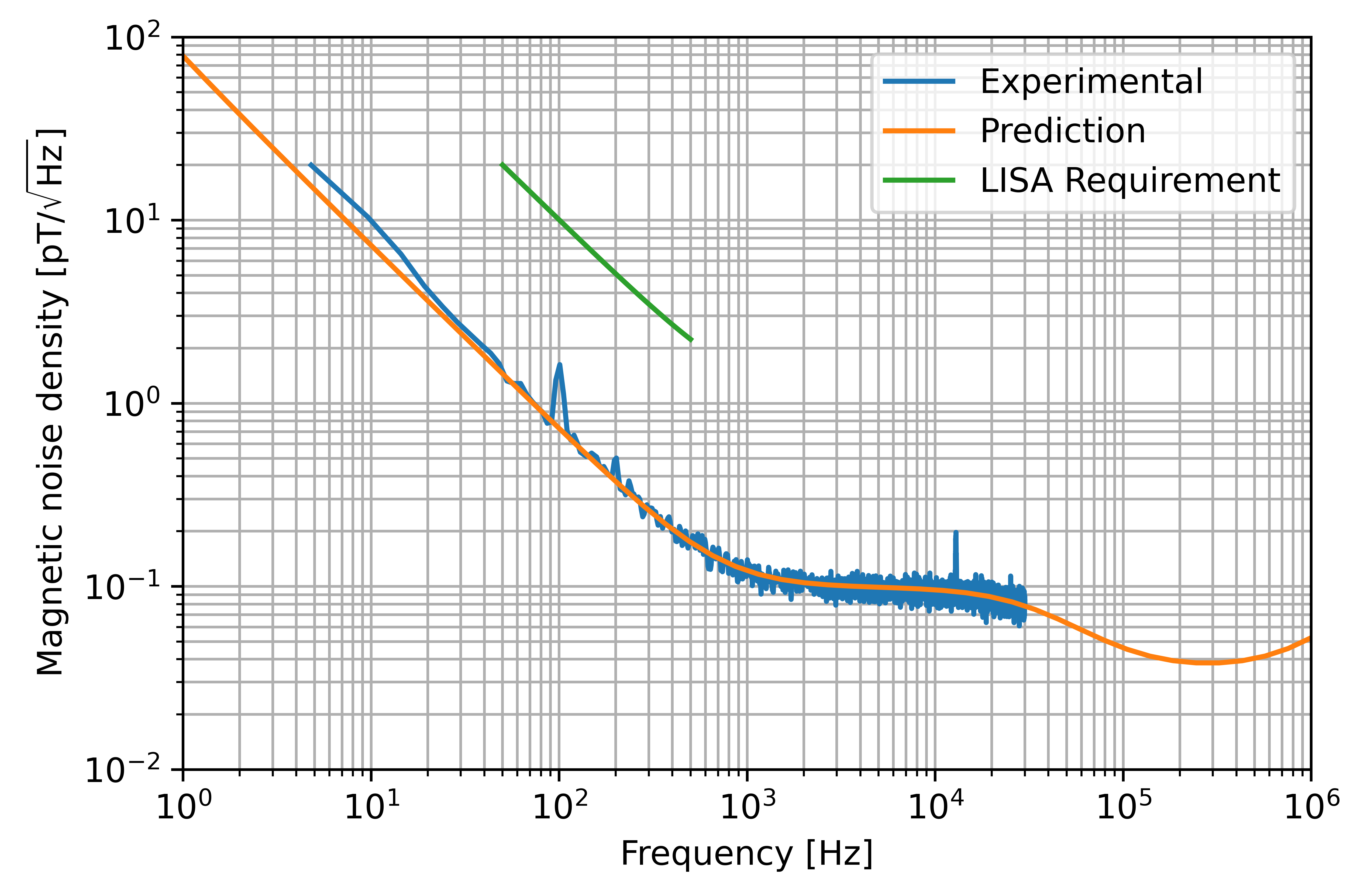}
    \caption{LPF coil's magnetic noise density experimental results shown alongside the predicted curve and the LISA requirement.}
    \label{fig:magnetic_with_data}
\end{figure}

These results can be converted into equivalent magnetic field noise by using the sensitivity from~(\ref{eq:B_noise}). Figure~\ref{fig:magnetic_with_data} shows the magnetic noise level performance of the LPF coil when used as a SCS magnetometer. The measurements follow the predicted noise curve as expected from the voltage noise density, proving the capability of the LPF coils of measuring (1.49 $\pm$ 0.16) $\rm pT/\sqrt{\rm{Hz}}$ at 50 Hz and (0.1738 $\pm$ 0.0023) $\rm pT/\sqrt{\rm{Hz}}$ at 500 Hz.

Finally, we can perform an experimental assessment of the detectability of a potential high frequency disturbance in LISA. We experimentally study the signal generated in our antenna by a 50 pT tone at 50 Hz, i.e. in the lowest frequency range of the device. 
In order to achieve such stringent magnetic field amplitude we placed the coil inside the mumetal chamber which itself was placed in the center of the Helmholtz coil. Using a spectrum analyzer, model Stanford-SR780, and with the previous knowledge of the coil sensitivity we injected a sinusoidal signal through the Helmholtz coil at a frequency of 50 Hz. The analyzer was set to have an integration time of 5 seconds, similar to the expected during LISA operations. The injected voltage to the Helmholtz was reduced manually in order to decrease the measured peak amplitude by the spectrum analyzer until it was around the 50 pT, which, from~(\ref{eq:SCS_sensitivity}), corresponds to a voltage of $\approx$ 30 $\mu$V at the output of the amplifier. Comparing the peak value with respect to the noise level from the spectrum analyzer measurements, we found that the 50 pT signals had a SNR of 20.2 $\pm$ 1.4 which allows for a correct detection by the read-out in the satellite. 
\section{Detection Range}\label{sec:det_range}



We finally provide an assessment of source detectability in terms of its orientation with respect to the coil axis. To do so, we will assume that magnetic field sources are approximated by magnetic dipoles oscillating at a frequency $\omega$ and that the magnetic field is constant within the coil surface. In such a case, we can re-write Faraday's law as follows for an external magnetic field of the type shown in~(\ref{eq:B_coils_paper})

\begin{equation}
    V = HN(\pi a^{2})\cos\theta\left( |\mathbf{B}|\omega + |\mathbf{B_{p}}|\delta A \right),
    \label{eq:voltage_cos_dependence_all}
\end{equation}

where $H$ is the transfer function from~(\ref{eq:transfer_func_modulus}) and $\theta$ is the angle between $\mathbf{B}$ and the axis of the coil. At high frequencies and for small variations in time of $A(t)$ the second term can be neglected, resulting in an expected voltage within the SCS such that

\begin{equation}
    V = H\omega N(\pi a^{2})B_{\rm eff}\cos\theta,
    \label{eq:voltage_cos_dependence}
\end{equation}

with 

\begin{equation}
    B_{eff} = \frac{\mu_{0}p_{\rm eff}}{2\pi r^{3}},
    \label{eq:efective_field}
\end{equation}

where $p_{\rm eff} = |\mathbf{p_{\rm eff}}|$ is the magnitude of the dipole magnetic moment of the source. This is the case when the dipole moment is aligned with the direction vector to the coil, $\mathbf{r}$, producing the largest magnetic field amplitude possible. The value of $B_{\rm eff}$ we will be considering is the smallest possible magnetic field within the spacecraft, $B_{\rm eff} \rm \simeq 10$ pT as expected at a distance of about 0.3 m. Hence, the value of the effective dipole moment is $p_{\rm eff} = 50 \,\mu$Am$^{2}$. 
\begin{figure}
    \centering
    \includegraphics[width=\linewidth]{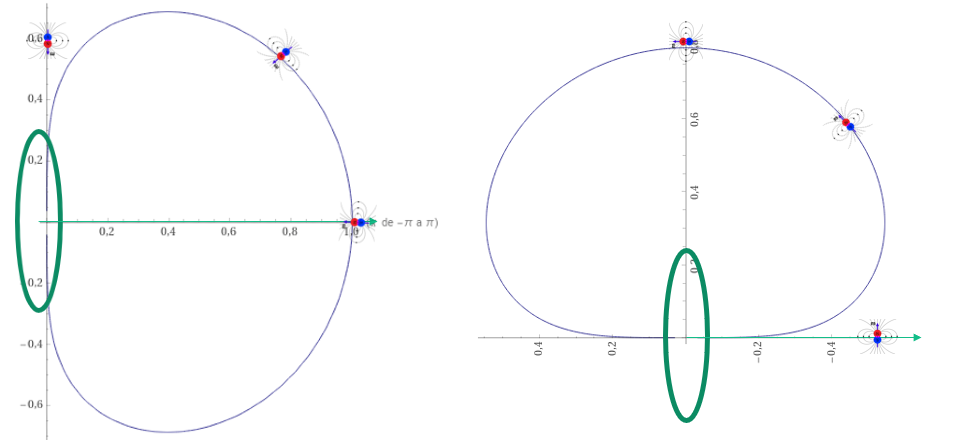}
    \caption{Polar graph of the function $r = \sqrt[3]{cos\theta}$ with the coil, not to scale, and the two detectable orientations of the magnetic dipoles. Left: $\overrightarrow{m}||\overrightarrow{r}$. Right: $\overrightarrow{m}\perp\overrightarrow{r}$ (smaller by $\sqrt[3]{2}$).}
    \label{fig:Polar_plot}
\end{figure}

We are interested in the case where they are orthogonal to each other, the amplitude of the magnetic field is then minimum and equal to $B_{\rm eff}/2$. Any other possible orientation of the magnetic dipole with respect to the coil distance vector can be expressed as a linear combination between the previous. Note that there are two perpendicular directions. Hence, there are a total of three orientations of the magnetic dipoles to be considered. However, there is one of the perpendicular orientations that will never induce a current into the coil because the induced field will always be orthogonal to the coil axis, $\theta = 90^{\circ}$. That is the reason why we will only be considering the two dipoles orientations within the plane of the image, as shown in figure~\ref{fig:Polar_plot}. More on this last possible dipole orientation later within the text.

In order to asses the detection range of the LPF coil, we will be analyzing the distance dependence with respect to the noise voltage of the SCS. From equation~(\ref{eq:voltage_cos_dependence}) we derive

\begin{equation}
    r = \sqrt[3]{H\omega N\pi a^{2}\frac{\mu_{0}p_{\rm eff}}{2\pi V}\cos\theta},
    \label{eq:distance_formula}
\end{equation}

which expresses the distance at which an effective dipole would induce a given voltage. To assess an envelope of detection we define a threshold given by the signal SNR, defined as ${\rm SNR} = V/V_{0}$, where $V$ is the voltage induced by the source and $V_{0}$ is the noise level voltage amplitude of the coil. The equivalent voltage noise assumed for this analysis is taken from figure~\ref{fig:voltage_with_data} to be a white floor noise of $S_{V}^{1/2}$ = 15 nV/$\sqrt{\rm{Hz}}$. This can be converted into voltage amplitude by using the normalized equivalent noise bandwidth of the Blackmann-Harris window, which leads to a voltage noise amplitude of $V_{0} \approx 35$ nV. 

\begin{figure}
    \centering
    \includegraphics[width=\linewidth]{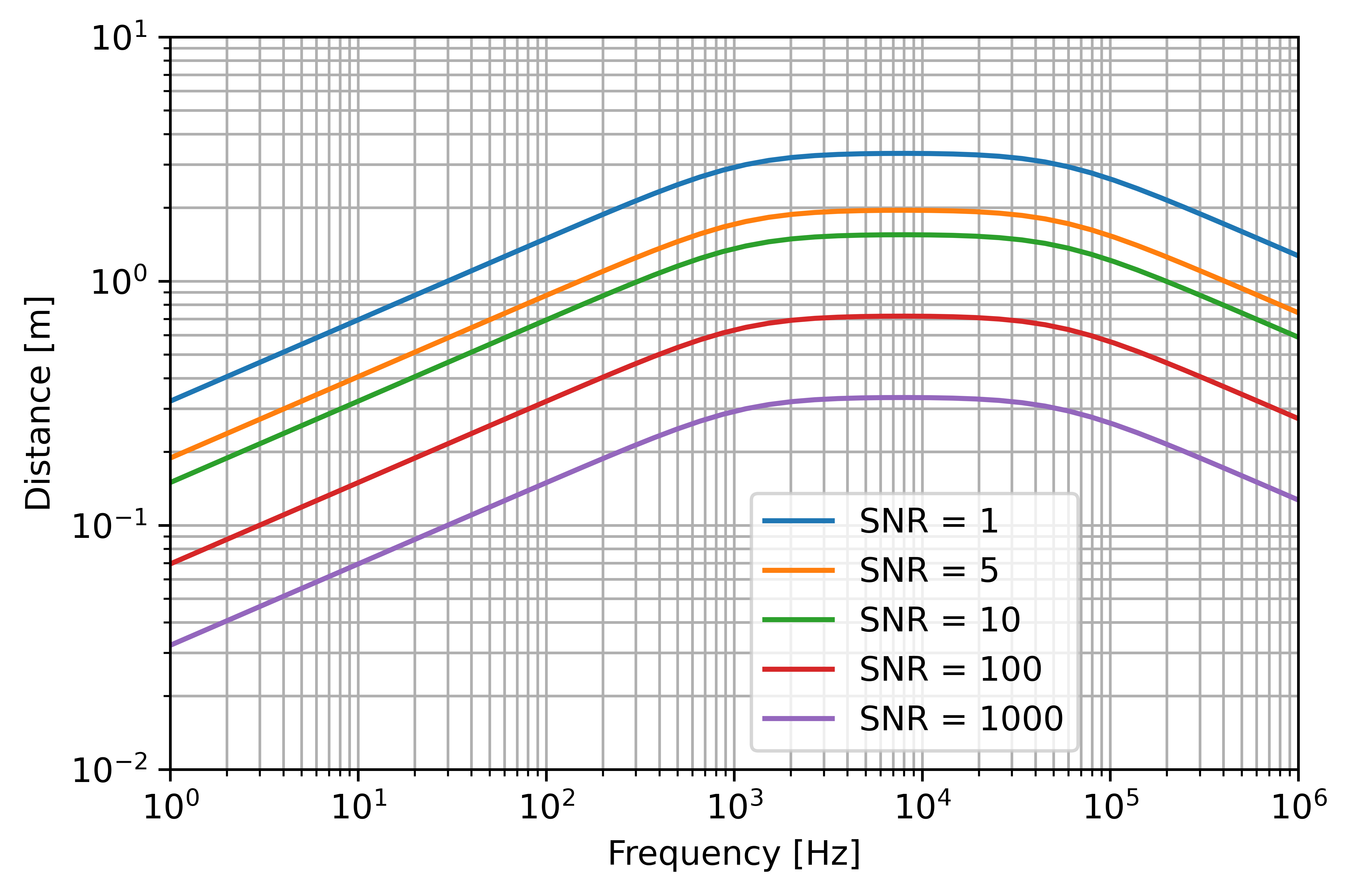}
    \caption{LPF coil detection range as a function of frequency and SNR.}
    \label{fig:SNR_with_distance}
\end{figure}

In figure~\ref{fig:SNR_with_distance} we show the detection range of the LPF SCS as a function of frequency. The effective detection distance increases for increasing frequencies and is reduced for higher SNR, as expected. The higher the SNR, the greater the required magnetic field amplitude, meaning the source must be closer to the coil. However, this detection range has a dependency on the angle of incidence between the magnetic field direction and the coil axis.



\begin{figure}
    \centering
    \includegraphics[width=0.8\linewidth]{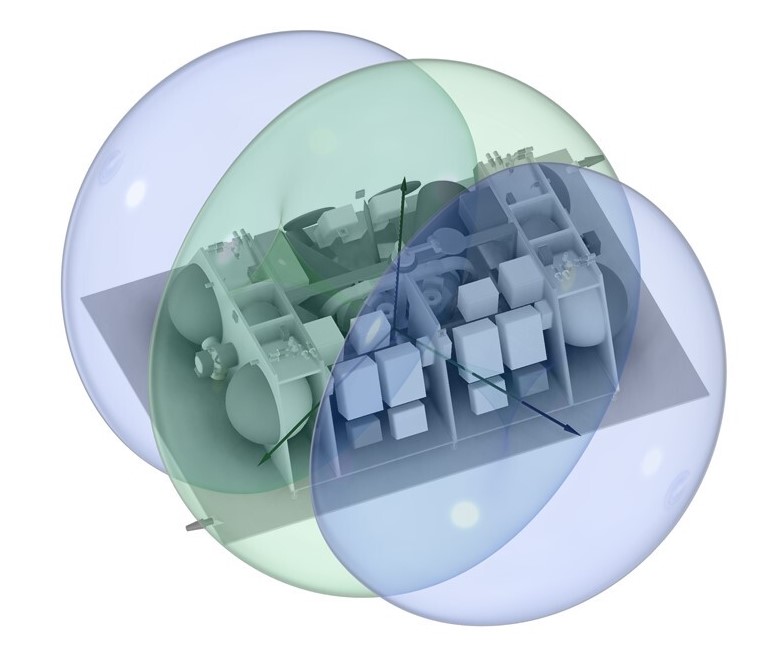}
    \caption{LPF coil detection lobes within the LISA spacecraft 3D model [Credit: ESA LISA Phase 0 Concurrent Design Facility study]. Blue: $\overrightarrow{m}||\overrightarrow{r}$. Green: $\overrightarrow{m}\perp\overrightarrow{r}$ (smaller by $\sqrt[3]{2}$).}
    \label{fig:DL}
\end{figure}

The polar distribution for the two types of orientations of $p_{\rm eff}$ is shown in figure~\ref{fig:Polar_plot}, as a function of the angle $\theta$ which has a $\sqrt[3]{\cos\theta}$ dependence from ~(\ref{eq:distance_formula}). As we can see, these are the detection lobes for the only two voltage inducing orientations of the magnetic dipole. On the left of the figure, we see the orientation of the magnetic dipole inducing the maximum amplitude magnetic field at the coil location ($\mathbf{m}$ and $\mathbf{r}$ are parallel). The blue line curve represents the detection lobe for dipole sources oriented in maximum configuration. On the right, we have the same plot but for the minimum magnetic field configuration ($\mathbf{m}$ and $\mathbf{r}$ are perpendicular).

The two dipole orientations can be measured by the coil and each have their own detection lobes. In figure~\ref{fig:DL}, we show the combination of both detection lobes together with a mock model of the LISA spacecraft and its units. The coils are located by the end of the telescopes (large cylinders within the image), one per telescope. In the figure we are only showing the detection lobe of a single coil in order to show their measurement potential. The black lines represent the axis of the coil and the one perpendicular to it, within the plane of the image. In green, we show the orthogonal orientations detection lobes and, in blue, we show the parallel dipole orientations detection lobes. The former is smaller by a factor $\sqrt[3]{2}$ because the amplitude of the magnetic field is smaller by a factor of 2. They have been built using the middle value within the audio frequency band, from figure~\ref{fig:SNR_with_distance} with a SNR of 10. 

As we can see, the coils will be able to cover the entire spacecraft with a high enough SNR even after assuming the smallest possible magnetic fields. Any larger signals will be measured with higher SNRs. One coil alone would be blind to specific orientations, such as the situation described previously or even the ones outside the detection lobes for their respective dipole orientations. For example, dipoles with $\mathbf{m}\parallel\mathbf{r}$ would not be observed outside the blue detection lobes. That's where the importance of including a second coil comes in. The other coil, with their current disposition within the spacecraft, would be able to detect the signals that the other coil is blind to. This is thanks to the orientation of the telescope within the spacecraft, making each coils axis approximately 60$^{\circ}$ from one another due to the LISA equilateral arm formation. If the coils were to be placed somewhere else in the final spacecraft design, this is something that should be accounted for.

Lastly, we wanted to note that the second orthogonal orientation of the dipoles would never be measured by any of the two coils. From figure~\ref{fig:DL}, the dipole orientation that is not measurable is the one aligned with the vertical axis of the spacecraft, perpendicular to the plane. That orientation always induce into both coils magnetic fields perpendicular to their surfaces. Despite the coils being blind to this last possible dipole orientation, the magnetic force such sources would induce to the TM would also be along the vertical axis. For the specific case of LISA, the forces that are of uttermost importance to monitor are the ones along the direction between spacecrafts (following the telescopes directions). This means that not being able to detect such orientations will not be critical for the mission. 



\section{Conclusions} \label{sec:conclusions}

We have presented a complete analysis of the performance of the LPF coil used as an audio-band frequency magnetometer for the LISA mission. As for the current LISA requirements, the LPF coil has been proven to perform an order of magnitude below the requirements set within the 50 - 500 Hz frequency band. The performance of the magnetic noise level of this sensor has been shown to be capable of measuring 1.45 $\rm pT/\sqrt{\rm{Hz}}$ at 50 Hz and 0.17 $\rm pT/\sqrt{\rm{Hz}}$ at 500 Hz. Additionally, beyond requirements, the LPF coil is able to measure a magnetic noise level as low as 0.1 $\rm pT/ \sqrt{Hz}$ within the kHz frequency band. With the design proposed in this work, the LPF coil used as a search coil sensor magnetometer can detect 50 pT magnetic signals at 50 Hz with a signal-to-noise ratio of at least 20. Additionally, the actually proposed distribution of the coils within the LISA spacecrafts allows the detection of all sources of high-frequency magnetic signals originated from the electronics on-board, which can affect the performance of the mission, with a high SNR.


\section*{Acknowledgements} 

This work has been made possible by the support from the Spanish Ministry of Science and Innovation No. PID 2022-137674NB- I00 and by the programme {\em Unidad de Excelencia Mar\'{\i}a de Maeztu} CEX2020-001058-M. Support from AGAUR (Generalitat de Catalunya) Contract No. 2021-SGR-01529 is also acknowledged. Thanks to Jonan Larranaga for the LISA spacecraft model from the ESA LISA Phase 0 Concurrent Design facility study.

\appendix


\bibliographystyle{iopart-num}    
\bibliography{library.bib}

\end{document}